\documentclass[10pt]{article}
\usepackage[total={6in,9.5in}]{geometry}
\usepackage{amsmath,amssymb,amsthm}
\usepackage{siunitx}
\usepackage{optidef}
\usepackage{palatino}
\usepackage{graphicx,epstopdf,epsfig}
\usepackage[dvipsnames]{xcolor}
\usepackage{float}
\usepackage[caption=false,font=footnotesize]{subfig}
\usepackage{url}
\usepackage{microtype}
\usepackage[normalem]{ulem}
\usepackage[section]{placeins} 
\usepackage{centernot}
\usepackage{enumitem}
\usepackage[noadjust]{cite}
\usepackage{booktabs}
\usepackage{multirow}
\usepackage{algorithmic}
\usepackage{algorithm}
\usepackage[percent]{overpic}
\usepackage[colorlinks=true,
  linkcolor=Green,
  citecolor=Blue,
  urlcolor=BrickRed]{hyperref}

\def\b{\boldsymbol{b}}

\def\x{\boldsymbol{x}}

\def\A{\mathbf{A}}

\def\K{\mathbf{K}}
\def\Re{\mathbb{R}}
\def\N{\mathbb{N}}
\def\cH{\mathcal{H}}  
\def\cK{\mathcal{K}}  

\begin{document}

\title{Stabilizing RED using the Koopman Operator}

\date{}

\author{Shraddha~Chavan  and  Kunal~N.~Chaudhury
\thanks{
Both authors are with the Indian Institute of Science, Bengaluru 560012, India.
\texttt{cshraddha@iisc.ac.in}, \texttt{kunal@iisc.ac.in}. This work was supported by grant STR/2021/000011 from the ANRF, Government of India.
}
}

\maketitle

\begin{abstract}
The widely used RED (Regularization-by-Denoising) framework uses pretrained denoisers as implicit regularizers for model-based reconstruction. Although RED generally yields high-fidelity reconstructions, the use of black-box denoisers can sometimes lead to instability. In this letter, we propose a data-driven mechanism to stabilize RED using the Koopman operator, a classical tool for analyzing dynamical systems. Specifically, we use the operator to capture the local dynamics of RED in a low-dimensional feature space, and its spectral radius is used to detect instability and formulate an adaptive step-size rule that is model-agnostic, has modest overhead, and requires no retraining. We test this with several pretrained denoisers to demonstrate the effectiveness of the proposed Koopman stabilization.
\end{abstract}

\section{Introduction}
\label{intro}

The goal in image reconstruction is to recover a clean image $\boldsymbol{x} \in \Re^n$ from measurements 
\begin{equation}
    \label{eq:fm}
    \boldsymbol{b} = \mathbf{A} \boldsymbol{x} + \boldsymbol{\varepsilon} \in \Re^m,
\end{equation}
where $\mathbf{A} \in \Re^{m \times n}$ is the forward operator and $\boldsymbol{\varepsilon} \in \Re^m$ is additive noise~\cite{bouman2022foundations}. This formulation subsumes inverse problems such as deblurring, superresolution, MRI, and tomography. Image reconstruction is an extensively studied problem, with a rich literature encompassing variational approaches~\cite{rudin1992nonlinear,geman1986markov,vese2016variational}, deep-learning models~\cite{dong2014learning,guo_mambair_2025,liang_swinir_2021,zamir2022restormer,kawar_denoising_2022,zhu_denoising_2023}, and denoiser-driven regularization~\cite {sreehari2016plug,hurault2022gradient,romano2017little,reehorst2018regularization,zhang2021plug,yuan2020plug}.

In model-based reconstruction, one typically considers the optimization problem
\begin{equation}
\label{eq:fplusg}
\min_{\x \in \Re^n} \ f(\x) + g(\x), \quad f(\x) = \frac{1}{2} \lVert \A\x - \b \rVert^2
\end{equation}
where $f$ enforces data fidelity and the regularizer $g : \Re^n \to [0, \infty]$ promotes image properties~\cite{vese2016variational,bouman2022foundations}. Traditional regularizers like total variational and wavelet sparsity~\cite{vese2016variational} often fail to capture rich structures in natural images. The key insight in~\cite{sreehari2016plug,romano2017little} was to replace hand-crafted regularizers with powerful off-the-shelf denoisers within classical iterative algorithms. This paper focuses on Regularization-by-Denoising (RED)~\cite{romano2017little}, in which the iterations are given by
\begin{equation}
\label{eq:red}
\x_{t+1} = \x_t - \gamma \Big( \nabla\! f(\x_t) +  \lambda \big( \x_t - D(\x_t)\big) \Big),
\end{equation}
where $\gamma$ is the step size, $\lambda$ is the regularization weight, and $D: \Re^n \to \Re^n$ is a denoising operator. As shown in~\cite{reehorst2018regularization}, under suitable assumptions on $D$, the update in~\eqref{eq:red} can be interpreted as gradient descent on an objective of the form~\eqref{eq:fplusg}.

Recent studies have shown that substantial improvements over traditional regularizers can be achieved when $D$ is a pretrained denoiser~\cite{cohen2021regularization,hurault2022gradient}. The key advantage of RED lies in the decoupling of the forward model and the denoiser in~\eqref{eq:red}, which allows a single pretrained denoiser to be applied across a wide range of imaging problems without retraining~\cite{romano2017little}. This flexibility, however, comes at the cost of theoretical guarantees—deep denoisers can lead to instability and degraded reconstructions~\cite{reehorst2018regularization,terris2024equivariant,nair2024averaged}. To address this issue, convergence-preserving denoisers have been proposed~\cite{hurault2022gradient,nair2024averaged,pesquet2021learning,hertrich_convolutional_2021,goujon_learning_2024}. A common approach is to interpret RED as a dynamical system,
\begin{equation}
\label{eq:redT}
\x_{t+1} = T_{\gamma}(\x_t), \quad T_{\gamma} = I - \gamma \big( \nabla f + \lambda (I - D) \big),
\end{equation}
and to establish convergence using properties of the operator $T_{\gamma}$, such as contractivity and averagedness~\cite{ryu2019plug,cohen2021regularization,nair2024averaged}. However, deep denoisers such as DnCNN and DRUNet do not satisfy the nonexpansiveness property~\cite{reehorst2018regularization,nair2024averaged}, which is necessary for establishing convergence guarantees. While structural constraints can enforce nonexpansiveness~\cite{ryu2019plug,pesquet2021learning,nair2023contractive,kumar2023lipschitz}, they can also degrade the regularization capacity.

\begin{figure*}[!t]
\centering
\setlength{\tabcolsep}{2.5pt}
\renewcommand{\arraystretch}{0}

\begin{tabular}{ccccc}
\multicolumn{1}{c}{$t=0$} & 
\multicolumn{1}{c}{$t=25$} &
\multicolumn{1}{c}{$t=200$} &
\multicolumn{1}{c}{$t=500$} &
\multicolumn{1}{c}{$t=1000$} \\
[0.3em]

    \begin{overpic}[width=0.19\textwidth]{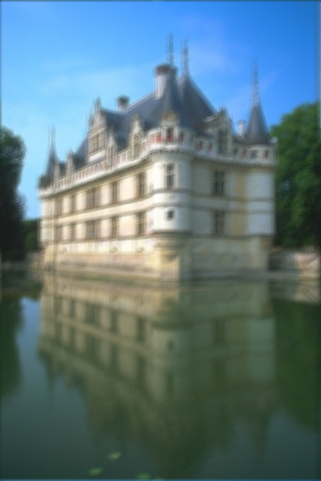}
        \put(3,93){\scriptsize\textbf{23.45}}
    \end{overpic} &
    \begin{overpic}[width=0.19\textwidth]{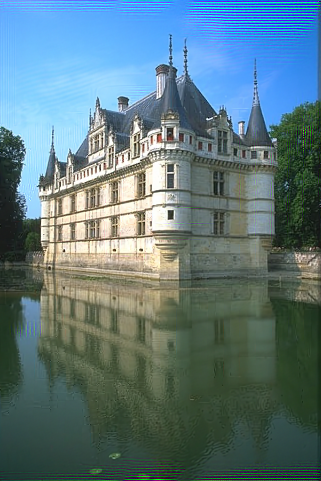}
        \put(3,93){\scriptsize\textbf{30.02}}
    \end{overpic} &
    \begin{overpic}[width=0.19\textwidth]{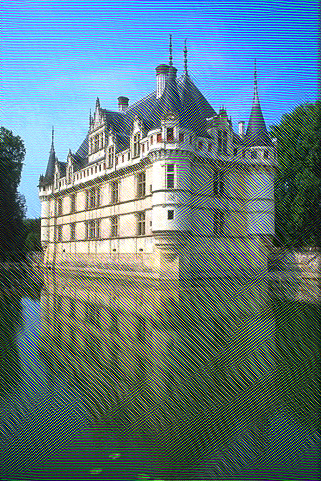}
        \put(3,93){\scriptsize\textbf{35.05}}
    \end{overpic} &
    \begin{overpic}[width=0.19\textwidth]{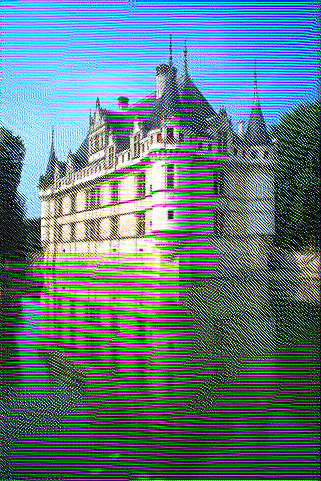}
        \put(3,93){\scriptsize\textbf{26.85}}
    \end{overpic} &
    \begin{overpic}[width=0.19\textwidth]{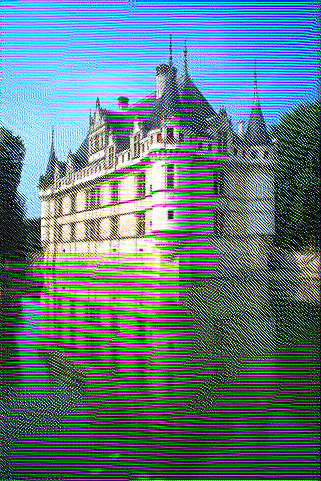}
        \put(3,93){\scriptsize\textbf{18.51}}
    \end{overpic}
    \\[0.5em]
    \begin{overpic}[width=0.19\textwidth]{images/RED_m2,n1_DRUNet/b_m,23.45.png}
        \put(3,93){\scriptsize\textbf{23.45}}
    \end{overpic} &
    \begin{overpic}[width=0.19\textwidth]{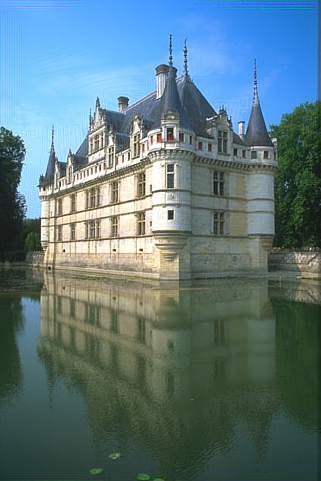}
        \put(3,93){\scriptsize\textbf{29.90}}
    \end{overpic} &
    \begin{overpic}[width=0.19\textwidth]{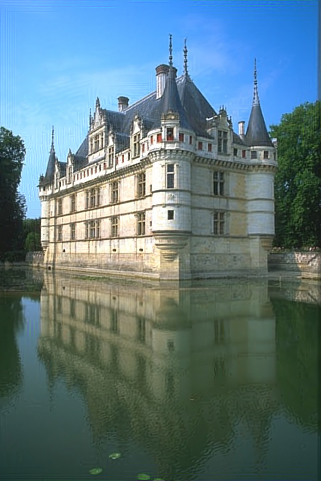}
        \put(3,93){\scriptsize\textbf{34.57}}
    \end{overpic} &
    \begin{overpic}[width=0.19\textwidth]{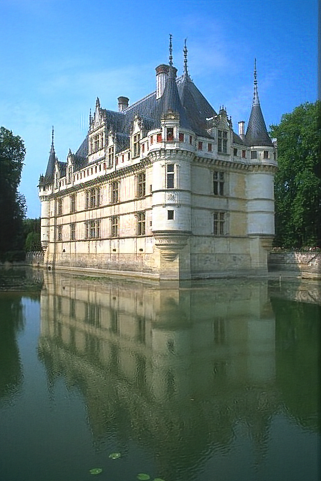}
        \put(3,93){\scriptsize\textbf{35.07}}
    \end{overpic} &
    \begin{overpic}[width=0.19\textwidth]{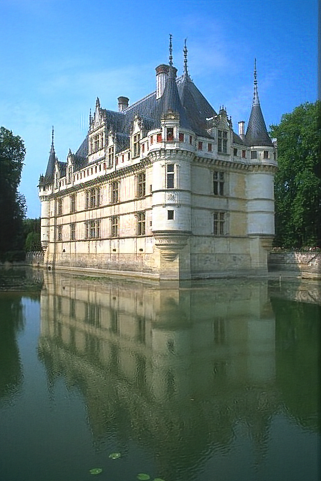}
        \put(3,93){\scriptsize\textbf{35.07}}
    \end{overpic}
\end{tabular}
\caption{
Results of motion deblurring with DRUNet using Vanilla RED (first row) and SKOOP-RED (second row). The PSNR value is shown at the top of each image.  We used the second motion kernel from~\cite{levin2009understanding}, and the noise level is $1/255$. While Vanilla RED shows early improvement, its performance starts degrading after a certain point, and artifacts appear across the image. In contrast, SKOOP-RED shows consistent improvement and stabilizes to a high-fidelity reconstruction. The corresponding quantitative trends are shown in Fig.~\ref{fig:motion_num}.}
\label{fig:progression_psnr}
\end{figure*}

We propose a stabilization mechanism for RED based on the classical Koopman operator~\cite{koopman1931hamiltonian,mezic2005spectral}. Instead of tracking RED iterates directly, we project them into a low-dimensional feature space and learn a Koopman operator $\K$ to model their dynamics~\cite{rowley2009spectral,brunton2017chaos}. By adaptively adjusting $\gamma$ in~\eqref{eq:red} using the spectral radius of $\K$, we achieve consistent stability improvements across pretrained denoisers, with less than $20\%$ overhead. To the best of our knowledge, this is the first work to leverage the Koopman operator for stabilizing iterative image reconstruction. 

The letter is organized as follows: the proposed stabilization is presented in Section~\ref{sec:SKOOP}; its effectiveness with deep denoisers is demonstrated in Section~\ref{sec:exp}; and we conclude in Section~\ref{sec:conclusion}. Additional experiments and ablation studies are provided in Section~\ref{sec:supplementary}.

\section{Koopman Stabilization}
\label{sec:SKOOP}

\subsection{Stability of RED}
\label{subsec:stability}

\begin{figure}[!t]
    \centering
    \includegraphics[width=\columnwidth]{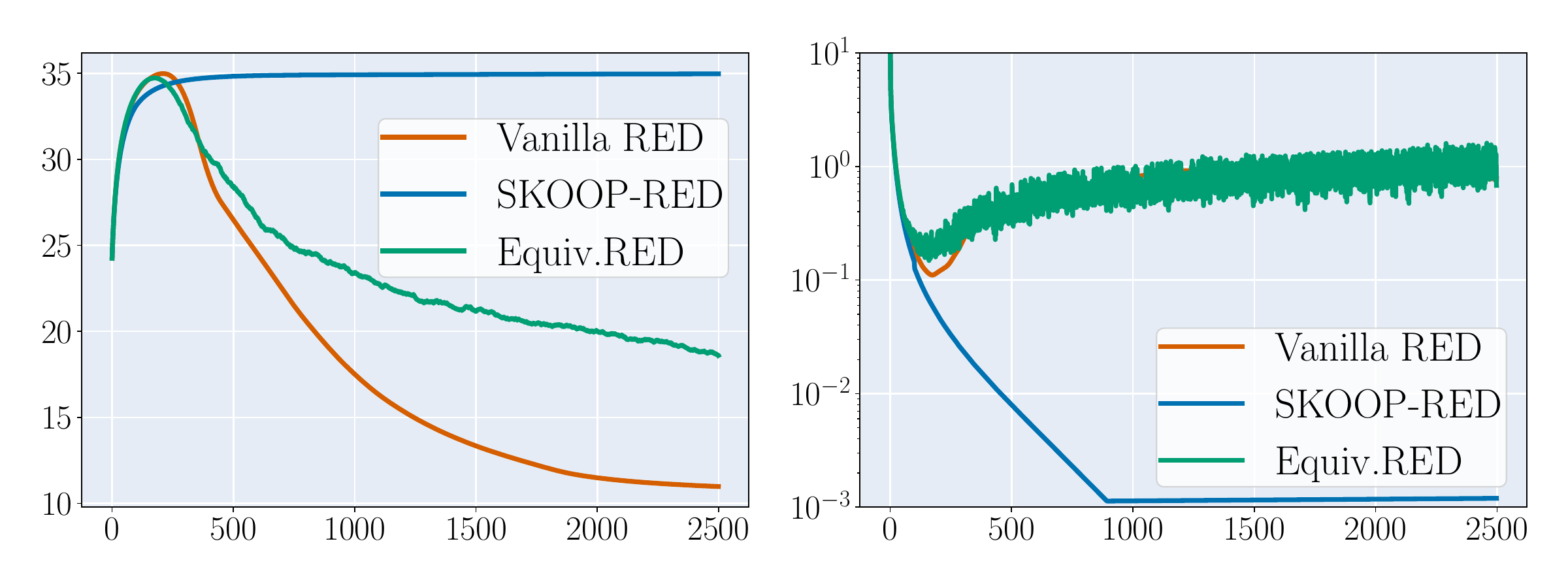}
   \caption{Results for motion deblurring
 using DRUNet denoiser. \textbf{Left:} PSNR (dB) vs.\ iteration. \textbf{Right:} residual norm $\lVert x_t - x_{t-1} \rVert$ vs.\ iteration $t$. Fig.~\ref{fig:progression_psnr} shows progressive reconstructions across iterations.}
    \label{fig:motion_num}
\end{figure}

\begin{figure}[!t]
    \centering
    \includegraphics[width=\columnwidth]{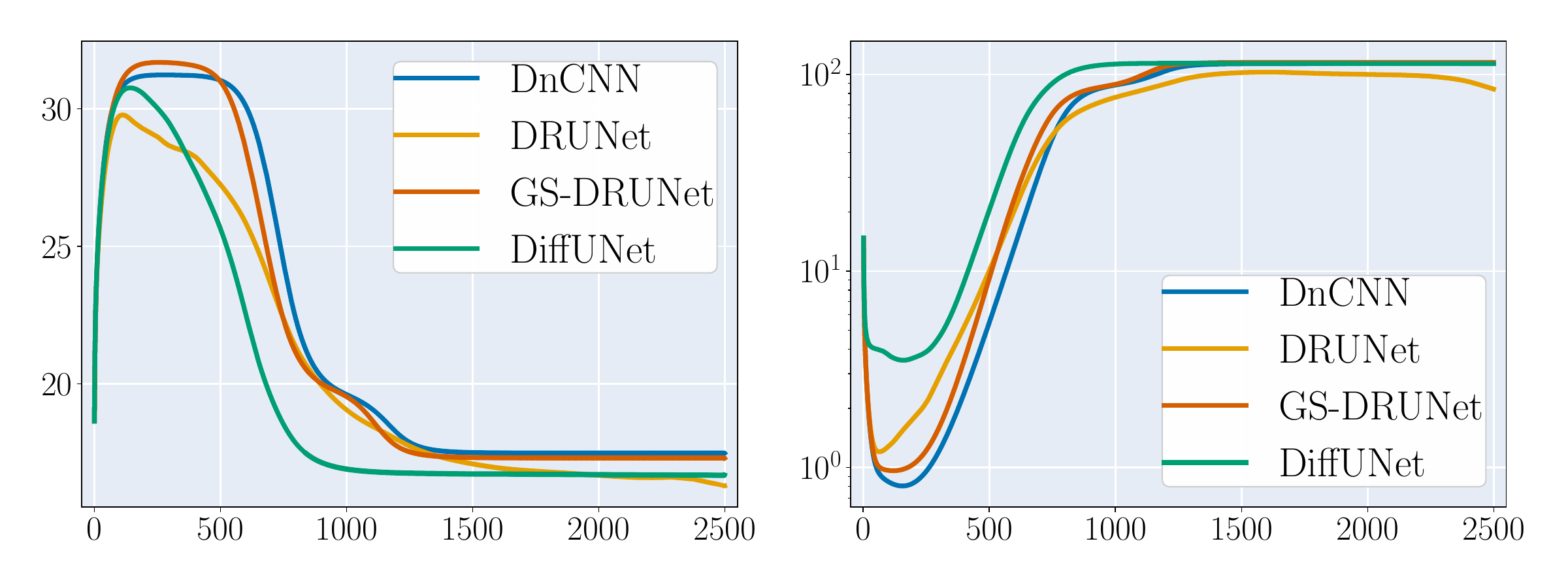}
    \caption{Instability of Vanilla RED for motion deblurring using four denoisers (DnCNN, DRUNet, GS-DRUNet, DiffUNet). \textbf{Left:} PSNR (dB) vs.\ iteration. \textbf{Right:} residual norm $\lVert x_t - x_{t-1} \rVert$ vs.\ iteration $t$. In all cases, PSNR eventually diverges sharply or gradually, regardless of the denoiser. Moreover, the breakdown point is unpredictable, making early stopping unreliable.}
    \label{fig:vanilla1}
\end{figure}

Unlike standard ML pipelines that apply a neural network once at inference, RED applies a denoiser repeatedly, which can lead to instability and cause the PSNR to diverge. As shown in Fig.~\ref{fig:vanilla1}, the PSNR improves initially but eventually degrades as instability sets in, a trend observed across different deep denoisers.  It was shown in~\cite{terris2024equivariant} that averaging over random transformations can mitigate instability to some extent. However, this approach merely delays the onset of divergence and does not eliminate it entirely~(see Fig.~\ref{fig:motion_num}).

\begin{algorithm}[!t]

\caption{SKOOP-RED}
\label{alg:SKOOP-red}
\begin{algorithmic}[1]
    \STATE \textbf{Input:} checkpoints $\Omega \subset \mathbb{N}$ and parameter $\beta >0$.
    \STATE \textbf{Initialization:}  $\boldsymbol{x}_0 \in \Re^n$,  $\gamma_0 > 0$
    \FOR {$t = 0, 1, \ldots$}
        \IF {$t \in \Omega$}
             \STATE Compute $\K_t$ using~\eqref{eq:koopman_dmd2}.
             \IF {$\varrho(\mathbf{K}_{t}) > 1$}
            \STATE Compute $\eta_t$ using~\eqref{eq:eta}. 
            \STATE  $\gamma_{t+1} = \eta_t \gamma_t$.
             \ENDIF
        \ELSE
            \STATE $\gamma_{t+1} = \gamma_t$.
        \ENDIF
        \STATE Update $\boldsymbol{x}_{t+1} = T_{\gamma_{t+1}}(\boldsymbol{x}_t)$ using~\eqref{eq:redT}.
    \ENDFOR
\end{algorithmic}
\end{algorithm}

While early stopping can help, the point where performance levels off depends heavily on the image, denoiser, and forward model, making it hard to identify. Analyzing the Jacobian of the RED update~\eqref{eq:red} is another option, but it captures only local behavior and misses the nonlinear dynamics. Moreover, computing and storing the Jacobian is costly, especially for large-scale problems. This motivates the need for a reliable and scalable method to stabilize the dynamical system~\eqref{eq:redT}.

\subsection{Koopman Operator}
\label{subsec:koopman}

The Koopman operator was introduced to bring Hilbert space and spectral analysis techniques to the study of dynamical systems~\cite{koopman1931hamiltonian}. Koopman's seminal insight was to identify a nonlinear evolution operator $T$ with a linear operator $\cK$ acting on a Hilbert space $\cH$ of \textit{observables} $\psi: \Re^n \to \mathbb{C}$. Specifically, the operator is defined as
\begin{equation}
    \label{eq:koopman}
    (\cK \psi)(\x) = \psi(T(\x)) \qquad (\psi \in \cH).
\end{equation}
In other words, \(\cK\) maps an observable $\psi$ to \(\cK \psi\), whose value at \(\x\) equals $\psi$ at the next state $T(\x)$ under the dynamics. The key observation is that although $T$ may be highly nonlinear, its Koopman operator $\mathcal{K}$ is always linear. This connection makes it possible to leverage the spectral theory of operators on Hilbert spaces~\cite{neumann1932proof}. 

The Koopman operator offers a principled framework for analyzing nonlinear dynamics, enabling insights into mixing, stability, and transients~\cite{mezic2004comparison,mezic2005spectral,rowley2009spectral,brunton2017chaos}. Its practical utility was advanced by dynamic mode decomposition (DMD)~\cite{schmid2010dynamic}, which approximates the infinite-dimensional operator with a finite linear model learned from trajectory data. This forms the basis of our stabilization mechanism.

\subsection{Proposed Stabilization}
\label{subsec:SKOOP-RED}

\begin{figure*}[!t]
    \centering
    \setlength{\tabcolsep}{2.5pt}
    \renewcommand{\arraystretch}{0}
    \begin{tabular}{cccc}
        \begin{overpic}[width=0.24\textwidth]{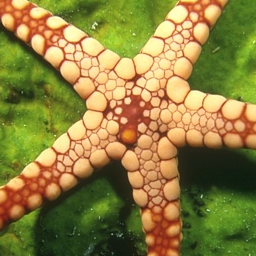}
            \put(30,93){\scriptsize\textbf{(a)~GT}}
        \end{overpic} &
        \begin{overpic}[width=0.24\textwidth]{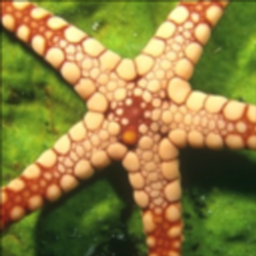}
            \put(30,93){\scriptsize\textbf{(b)~22.92~dB}}
        \end{overpic} &
        \begin{overpic}[width=0.24\textwidth]{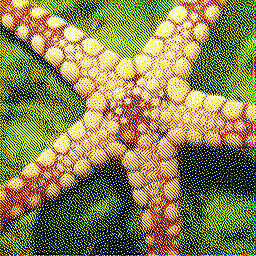}
            \put(30,93){\scriptsize\textbf{(c)~8.51~dB}}
        \end{overpic} &
        \begin{overpic}[width=0.24\textwidth]{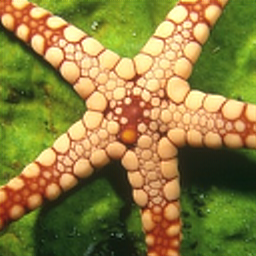}
            \put(30,93){\scriptsize\textbf{(d)~30.22~dB}}
        \end{overpic}
    \end{tabular}
    
    \caption{Results for $2\times$ superresolution using the DiffUNet denoiser, where the clean image is blurred with a $25{\times}25$ Gaussian kernel ($\sigma=1$) and downsampled. (a) Ground truth, (b) observed image, (c) reconstruction via Vanilla RED, and (d) reconstruction via SKOOP-RED. Both methods are run for $2500$ iterations, as detailed in Fig.~\ref{fig:sr_num}. SKOOP-RED produces stable, high-quality reconstructions, avoiding the instability seen in Vanilla RED.}
\label{fig:sr_diffunet_results}
\end{figure*}

\begin{figure}[!t]
    \centering
    \includegraphics[width=0.8\columnwidth]{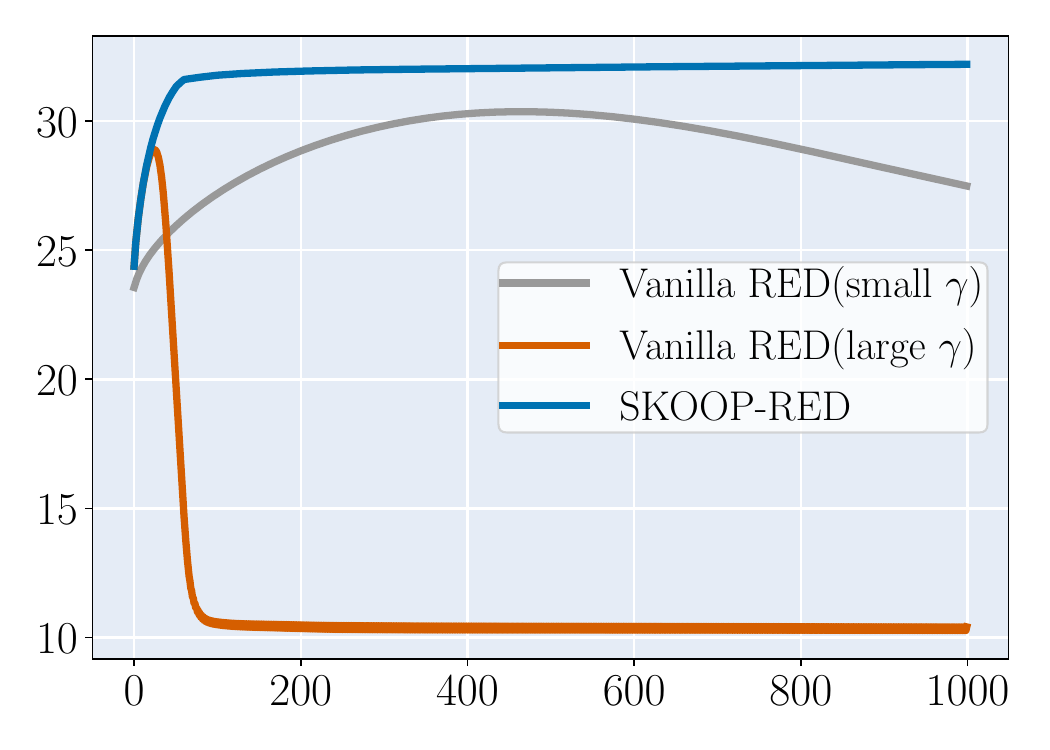}
    \caption{Comparison of SKOOP-RED with manually tuned step-size schedules. PSNR (dB) vs. iteration shows that SKOOP-RED consistently produces stable, high-quality reconstructions, whereas fixed decay strategies often stagnate or become unstable. This underscores the importance of adaptive step-size selection in SKOOP-RED.}
    \label{fig:gamma-motivation}
\end{figure}

Classical RED algorithms—RED-SD, RED-ADMM, RED-FP~\cite{romano2017little}—and their extensions, RED-APG, RED-DPG, RED-PRO~\cite{reehorst2018regularization,cohen2021regularization}, use a fixed, manually-tuned step size $\gamma$. For instance, RED-SD, which we refer to as \textit{vanilla} RED in this paper, uses a fixed $\gamma$ derived from technical assumptions on the denoiser. However, it was observed in~\cite{reehorst2018regularization} that deep denoisers tend to violate these conditions.  As shown in Fig.~\ref{fig:gamma-motivation}, reconstruction quality suffers when $\gamma$ is too small or too large. Notably, a large $\gamma$ can lead to instability after a few iterations. 

We will show that data-driven step size schedules can significantly improve the stability of RED. Our approach, called SKOOP (\textbf{S}tabilization using the \textbf{Ko}opman \textbf{Op}erator), adjusts the step size dynamically based on the spectral radius of the Koopman operator. This is outlined in Algorithm~\ref{alg:SKOOP-red}.

The step size correction in SKOOP-RED is performed at checkpoints $\Omega \subset \N$, balancing computational efficiency with effective monitoring. For each checkpoint $t \in \Omega$, we consider a history of past samples:
\begin{equation*}
X_t= \Big\{\x_t : t-w+1 \leqslant \tau \leqslant t \Big\} \subset \Re^n.
\end{equation*}

We project $X_t$ into a low-dimensional space using a feature map $\psi: \Re^n \to \Re^d, \, d \ll n$, analogous to an observable in the Koopman formulation~\eqref{eq:koopman}. We next learn a linear operator $\K_t \in \Re^{d \times d}$ such that for $t-w+1 \leqslant \tau \leqslant t$,
\begin{equation}
\label{eq:koopman_dmd1}
\K_{t} \psi(\x_{\tau}) = \psi(T(\x_{\tau})) = \psi(\x_{\tau+1}).
\end{equation}
Thus, similar to~\eqref{eq:koopman}, $\K_t$ progresses the dynamical system (in the feature space) by one time step. However, such an operator need not always exist. Therefore, following~\cite{schmid2010dynamic, rowley2009spectral}, we estimate $\K_t$ using linear regression:
\begin{equation}
\label{eq:koopman_dmd2}
\K_t= \underset{\K}{\arg \min} \sum_{\tau =t-w+1}^{t-1} \, \big\| \K\, \psi(\x_{\tau})  - \psi(\x_{\tau+1}) \big\|^2.
\end{equation}
This is solvable using efficient numerical techniques~\cite{tu2013dynamic, Korda_2017}. 

The spectral radius $\varrho(\K_t)$ is used as an indicator for local stability in the feature space: $\varrho(\K_t) < 1$ implies decaying dynamics in the feature space, while $\varrho(\K_t) \geqslant 1$ indicates potential instability. In the latter case, we shrink the step size by a factor
\begin{equation}
\label{eq:eta}
    \eta_t = \exp\left(-\beta (\varrho(\mathbf{K}_t) - 1)\right),
\end{equation}
where $\beta > 0$ controls the rate of decay. We set $\beta = 2$ based on empirical validation, as this choice consistently provided a favorable balance between stability and efficiency across all tested scenarios. Larger values of $\beta$ result in rapid decay in the step size, hindering progress, while smaller values are less effective at mitigating instability, particularly in more challenging cases. This completes the description of SKOOP-RED. We note that the operator $T_{\gamma}$ in Algorithm~\ref{alg:SKOOP-red} is the Vanilla RED update~\eqref{eq:redT}.

\begin{table}[!t]

\caption{
Per-iteration runtime (sec) and overhead relative to Vanilla RED (\%). The results are averaged over the Set15C dataset and imaging tasks.
}
\vspace{0.4em}
\centering
\setlength{\tabcolsep}{6pt}
\renewcommand{\arraystretch}{1.1}
\begin{tabular}{|l|c|c|c|c|c|}
\hline
\multirow{2}{*}{\textbf{denoiser}} & \multicolumn{3}{c|}{\textbf{run time (sec)}} & \multicolumn{2}{c|}{\textbf{overhead (\%)}} \\
\cline{2-6}
 & Vanilla & Equivariant & SKOOP & Equivariant & SKOOP \\
\hline
DnCNN    & 0.0163 & 0.0166 & 0.0186 & 1.84 & \textbf{14.11} \\
DRUNet   & 0.0214 & 0.0217 & 0.0240 & 1.38 & \textbf{12.15} \\
GS-DRUNet & 0.0238 & 0.0242 & 0.0265 & 1.68 & \textbf{11.34} \\
DiffUNet & 0.0383 & 0.0388 & 0.0425 & 1.31 & \textbf{10.97} \\
\hline
\end{tabular}
\label{tab:runtime_overhead}
\end{table}

\subsection{Implementation}

We explored various feature maps $\boldsymbol{\psi}$ to balance reliability and efficiency, and ultimately selected a $d=66$ dimensional descriptor combining per-channel global statistics, $4{\times}4$ spatial pooling, and four low-frequency DCT coefficients. Snapshots were collected over a window $w \in [30, 40]$ with checkpoints at $\Omega = \{ w + t r : t \geqslant 1 \}$, using a stride $r = 10$.  The above choices were validated through sensitivity studies and generalized well across architectures and forward models (see Section~\ref{sec:supplementary}).

We compute $\mathbf{K}_t$ using the technique in~\cite{schmid2010dynamic}, and evaluate its spectral radius $\varrho(\mathbf{K}_t)$ using an efficient linear algebra routine. For the configuration $d = 66, w = 30$, and $r = 10$, our implementation requires approximately $42$ milliseconds per iteration to perform motion deblurring on a $256 \times 256$ color image using the DiffUNet denoiser. This is an overhead of $11\%$ compared to Vanilla RED ($38.35$ ms). Of this total runtime, denoising accounted for $6\%$, feature extraction for $4.4\%$, and the Koopman operator and spectral radius computation for less than $0.1\%$. The overall runtime overhead relative to Vanilla RED remains consistently below $20\%$.

\section{Experiments}
\label{sec:exp}

\begin{table*}[!t]
\caption{
Peak/final PSNR values averaged over Set15C for motion deblurring across $8$ motion blur kernels~\cite{levin2009understanding}. By final PSNR, we mean the PSNR after $1000$ iterations. The baselines are Vanilla RED and Equivariant RED, and the SKOOP results are in bold.
}
\vspace{0.4em}
\centering
\setlength{\tabcolsep}{2.5pt}
\renewcommand{\arraystretch}{1.1}

\begin{tabular}{|c|ccc|ccc|}
\hline
\textbf{kernel}
& \multicolumn{3}{c|}{\textbf{DnCNN}}
& \multicolumn{3}{c|}{\textbf{DRUNet}} \\
\cline{2-7}
 & \textbf{Vanilla} & \textbf{Equiv.} & \textbf{SKOOP}
 & \textbf{Vanilla} & \textbf{Equiv.} & \textbf{SKOOP} \\
\hline
(a) & 28.01/10.79 & 28.02/10.80 & \textbf{34.36/34.36} & 28.05/10.76 & 28.06/10.77 & \textbf{34.65/34.65} \\
(b) & 26.67/10.75 & 26.67/10.75 & \textbf{33.44/33.44} & 26.83/10.72 & 26.84/10.73 & \textbf{34.56/34.56} \\
(c) & 29.60/10.88 & 29.62/10.87 & \textbf{33.12/33.12} & 29.97/10.74 & 29.98/10.82 & \textbf{34.32/34.32} \\
(d) & 27.93/11.11 & 27.95/11.11 & \textbf{33.02/33.02} & 28.08/10.97 & 28.10/11.02 & \textbf{33.62/33.62} \\
(e) & 29.55/11.05 & 29.56/11.06 & \textbf{34.88/34.88} & 29.68/10.93 & 29.68/10.96 & \textbf{35.85/35.85} \\
(f) & 28.25/11.12 & 28.26/11.13 & \textbf{35.49/35.49} & 28.35/11.09 & 28.36/11.10 & \textbf{36.49/36.49} \\
(g) & 28.70/10.73 & 28.73/10.77 & \textbf{32.82/32.82} & 28.93/10.57 & 28.94/10.67 & \textbf{34.36/34.36} \\
(h) & 27.22/10.97 & 27.24/10.97 & \textbf{32.30/32.29} & 27.28/10.73 & 27.29/10.83 & \textbf{32.74/32.74} \\
\hline
\end{tabular}

\vspace{0.4em}

\begin{tabular}{|c|ccc|ccc|}
\hline
\textbf{kernel}
& \multicolumn{3}{c|}{\textbf{GS-DRUNet}}
& \multicolumn{3}{c|}{\textbf{DiffUNet}} \\
\cline{2-7}
 & \textbf{Vanilla} & \textbf{Equiv.} & \textbf{SKOOP}
 & \textbf{Vanilla} & \textbf{Equiv.} & \textbf{SKOOP} \\
\hline
(a) & 28.04/10.78 & 28.05/10.78 & \textbf{35.27/35.27} & 28.08/10.79 & 28.08/10.79 & \textbf{35.02/35.02} \\
(b) & 26.85/10.75 & 26.86/10.75 & \textbf{34.79/34.79} & 26.89/10.75 & 26.89/10.75 & \textbf{34.91/34.91} \\
(c) & 30.10/10.88 & 30.11/10.87 & \textbf{34.86/34.85} & 30.15/10.87 & 30.16/10.86 & \textbf{34.94/34.94} \\
(d) & 28.11/11.11 & 28.12/11.11 & \textbf{34.27/34.27} & 28.17/11.09 & 28.17/11.09 & \textbf{34.13/34.13} \\
(e) & 29.74/11.04 & 29.75/11.04 & \textbf{36.10/36.10} & 29.77/11.03 & 29.77/11.03 & \textbf{36.21/36.20} \\
(f) & 28.38/11.13 & 28.38/11.13 & \textbf{36.83/36.83} & 28.41/11.12 & 28.41/11.12 & \textbf{36.96/36.96} \\
(g) & 29.02/10.74 & 29.03/10.72 & \textbf{34.88/34.87} & 29.05/10.71 & 29.05/10.72 & \textbf{35.05/35.04} \\
(h) & 27.35/10.97 & 27.37/10.97 & \textbf{33.82/33.82} & 27.37/10.96 & 27.37/10.96 & \textbf{33.64/33.63} \\
\hline
\end{tabular}

\label{tab:motion_psnr}
\end{table*}

\begin{figure}[!b]
    \centering
    \includegraphics[width=\columnwidth]{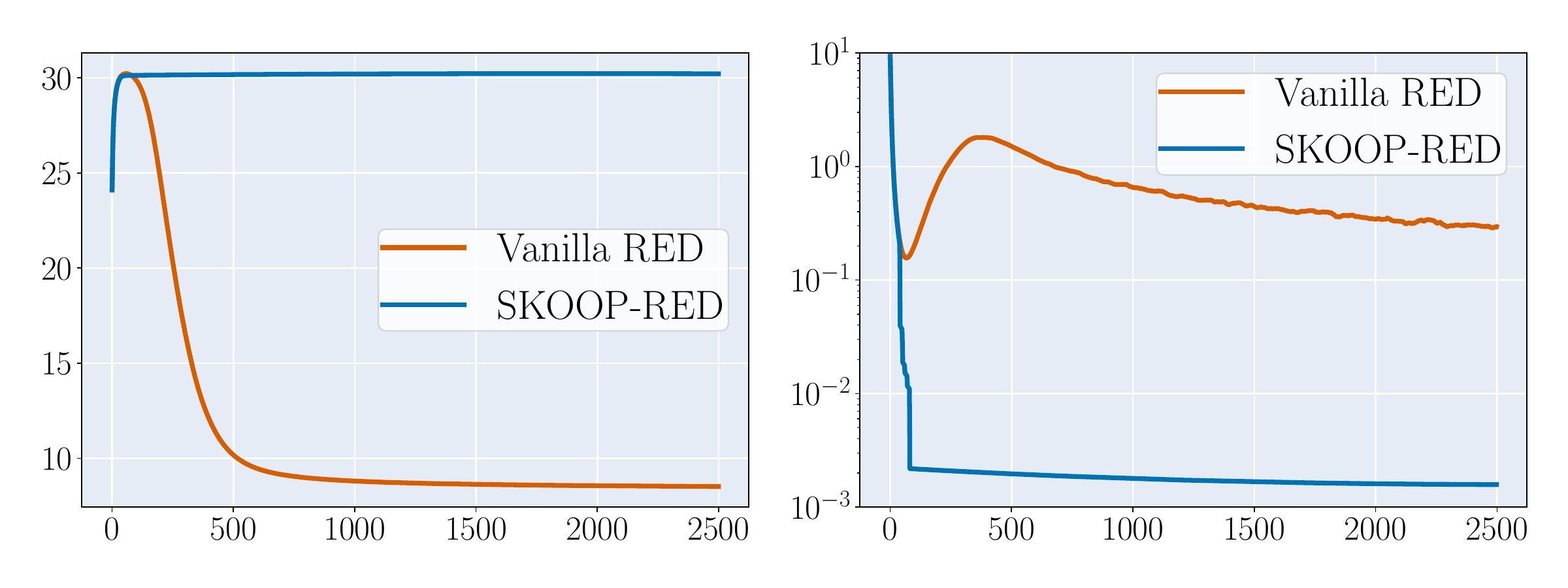}
    \caption{Results for $2\times$ superresolution ($25{\times}25$ Gaussian blur, $\sigma =1$) using DiffUNet denoiser. \textbf{Left:} PSNR (dB) vs.\ iteration. \textbf{Right:} residual norm $\lVert x_t - x_{t-1} \rVert$ vs.\ iteration $t$. See~Fig.~\ref{fig:sr_diffunet_results} for visual comparisons.}
  \label{fig:sr_num}
\end{figure}

We demonstrate the effectiveness of SKOOP-RED on three tasks: (i) Gaussian deblurring with blur levels $\sigma \in {1.0, 1.2, 1.6}$ and additive noise $\sigma_n \in {1/255, 2.55/255, 5/255}$, (ii) motion deblurring using benchmark kernels from~\cite{levin2009understanding}, and (iii) $2\times$ image superresolution. We initialize $\x_0$ with the observed image for deblurring and its bicubic interpolation for superresolution.

We experimented with four popular denoisers, DnCNN~\cite{zhang2017beyond}, DRUNet~\cite{zhang2021plug}, GS-DRUNet~\cite{hurault2022gradient}, and DiffUNet~\cite{ho2020denoising}, spanning both convolutional and attention-based architectures. We used the implementations in the DeepInverse library~\cite{tachella2025deepinverse}. The test set (Set15C) comprises $3$ images from Set3C and $12$ from BSD68. All experiments were performed on a system with an Intel Xeon Gold 6226R CPU, 502\,GB RAM, and an NVIDIA RTX A6000 GPU (48\,GB VRAM). The code is available at \url{https://github.com/Shraddha22710/SKOOP-RED}.

We compare our results with Vanilla RED~\cite{romano_red_repo, reehorst_repo} and equivariant RED~\cite{terris2024equivariant}. Methods such as group-averaging and DPIR-style self-ensembling~\cite{zhang2021plug} were excluded due to high cost. Figure~\ref{fig:progression_psnr} illustrates the PSNR and visual evolution during motion deblurring with DRUNet, supported by quantitative trends in Fig.~\ref{fig:motion_num}. Similarly, Fig.~\ref{fig:sr_diffunet_results} presents $2\times$ superresolution results using DiffUNet, with corresponding PSNR and residual norm plots in Fig.~\ref{fig:sr_num}. These highlight the frequent divergence of baseline methods, marked by PSNR drops, rising residual norms, and artifacts such as banding and grid patterns.

Table~\ref{tab:runtime_overhead} shows that SKOOP-RED adds only a small 15–20\% runtime overhead per iteration, thanks to efficient Koopman updates. As shown in Table~\ref{tab:motion_psnr}, it also achieves the most stable and highest final PSNR across all motion blur kernels and denoisers. In contrast, Vanilla RED often produces unstable results with visible artifacts, even when the PSNR is high (see Fig.~\ref{fig:progression_psnr}). This highlights that visual artifacts can appear long before any drop in metrics, making Vanilla RED unreliable for critical tasks. In contrast, SKOOP-RED consistently improves and preserves artifact-free reconstructions, which is essential for iterative image reconstruction. Additional visual and quantitative results are provided in Section~\ref{sec:supplementary}.

\section{Conclusion}
\label{sec:conclusion}

We demonstrated that the Koopman operator provides a simple yet effective data-driven mechanism for stabilizing RED with trained deep denoisers—a setting where obtaining convergence guarantees under verifiable assumptions remains elusive. In principle, the framework extends beyond RED-GD to other variants and shows promise for broader classes of iterative reconstruction methods~\cite{sreehari2016plug,ryu2019plug,zhang2021plug}. The Koopman viewpoint offers a fresh perspective on stabilization that should be of interest to the community.

\bibliographystyle{unsrt}

\clearpage
\appendix
\renewcommand{\thefigure}{S\arabic{figure}}
\setcounter{figure}{0}
\renewcommand{\thetable}{S\arabic{table}}
\setcounter{table}{0}

\section{Supplementary Material}
\label{sec:supplementary}

\begin{figure}[H]
    \begin{minipage}{\columnwidth}
        \centering
        \includegraphics[width=0.32\columnwidth]{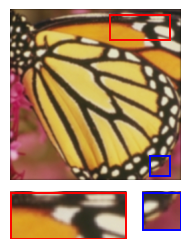}%
        \includegraphics[width=0.32\columnwidth]{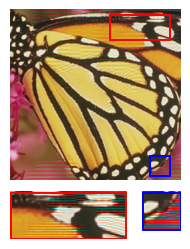}%
        \includegraphics[width=0.32\columnwidth]{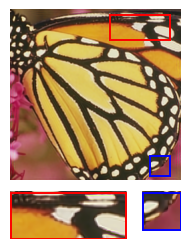}%
    \end{minipage}
    
    \centering
    \includegraphics[width=\columnwidth]{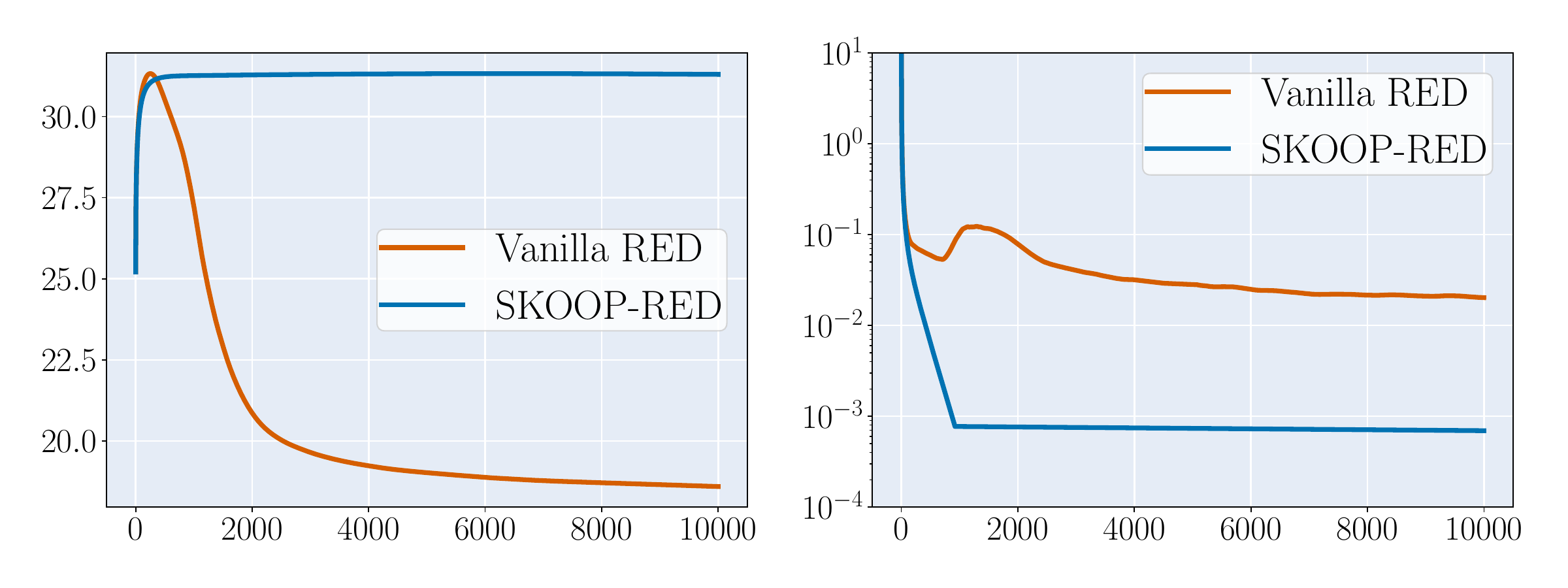}
     \caption{Example of Gaussian deblurring ($15 \times 15$, $\sigma = 1.2$) illustrating the long-term stability of SKOOP-RED with the DiffUNet denoiser. \textbf{Top:} Left to right—observed (23.36 dB), reconstruction via Vanilla RED (18.55 dB, SSIM 0.7735), reconstruction via SKOOP-RED (31.37 dB, SSIM 0.9554). \textbf{Bottom}: \textbf{Left:} PSNR (dB) vs.\ iteration. \textbf{Right:} residual norm $\lVert x_t - x_{t-1} \rVert$ vs.\ iteration $t$. SKOOP-RED remains stable over $10^4$ iterations and maintains peak PSNR, while Vanilla RED suffers a performance drop, as reflected in the declining PSNR curve.}
  \label{fig:longterm_stability}
\end{figure}

\subsection*{Feature map design for SKOOP-RED:}

The RED iterate $x_t$ is turned into a 66-dimensional feature vector at each Koopman checkpoint. This gives a compact but informative state description, allowing stable and efficient monitoring even when using black-box deep denoisers.
\begin{itemize}[leftmargin=*, itemsep=0pt]
    \item \textbf{Step-by-step construction:}
    \begin{enumerate}[label=(\alph*), itemsep=0pt]
        \item For each color channel:
        \begin{itemize}
            \item Compute the mean and standard deviation of all pixel values (2 features).
            \item Divide the channel into a non-overlapping $4\times4$ grid, and compute the mean in each cell (16 features).
            \item Apply the 2D discrete cosine transform (DCT) to the channel and extract the four lowest-frequency coefficients (top-left $2\times2$ block, 4 features).
        \end{itemize}
        \item Concatenate the 22 features from each channel to obtain $22 \times 3 = 66$ features for the image.
    \end{enumerate}
  \item This design provides a compact summary for robust, data-driven instability detection in RED. Empirically, global statistics capture large-scale deviations, spatial pooling detects local anomalies, and DCT features signal emerging instability, forming an efficient basis for stability monitoring.
\item The modular feature map enables Koopman-based real-time stability control and can be readily adapted to new observables, imaging tasks, or more advanced dynamics (see codebase for details).
\end{itemize}

\begin{figure}[!t]
    \centering
    \begin{minipage}{0.49\columnwidth}
        \centering
        \includegraphics[width=\linewidth]{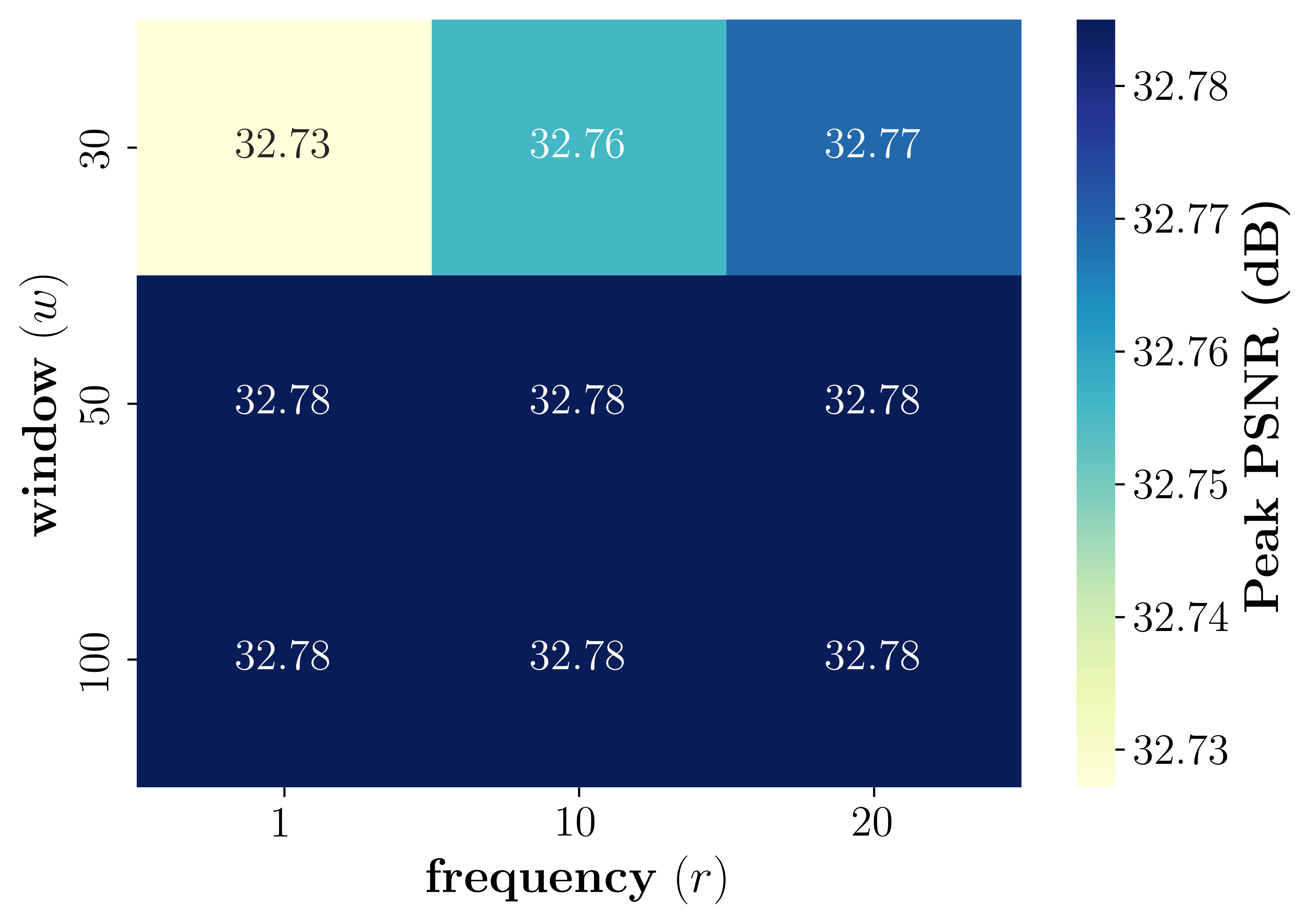}
    \end{minipage}
    \hfill
    \begin{minipage}{0.49\columnwidth}
        \centering
        \includegraphics[width=\linewidth]{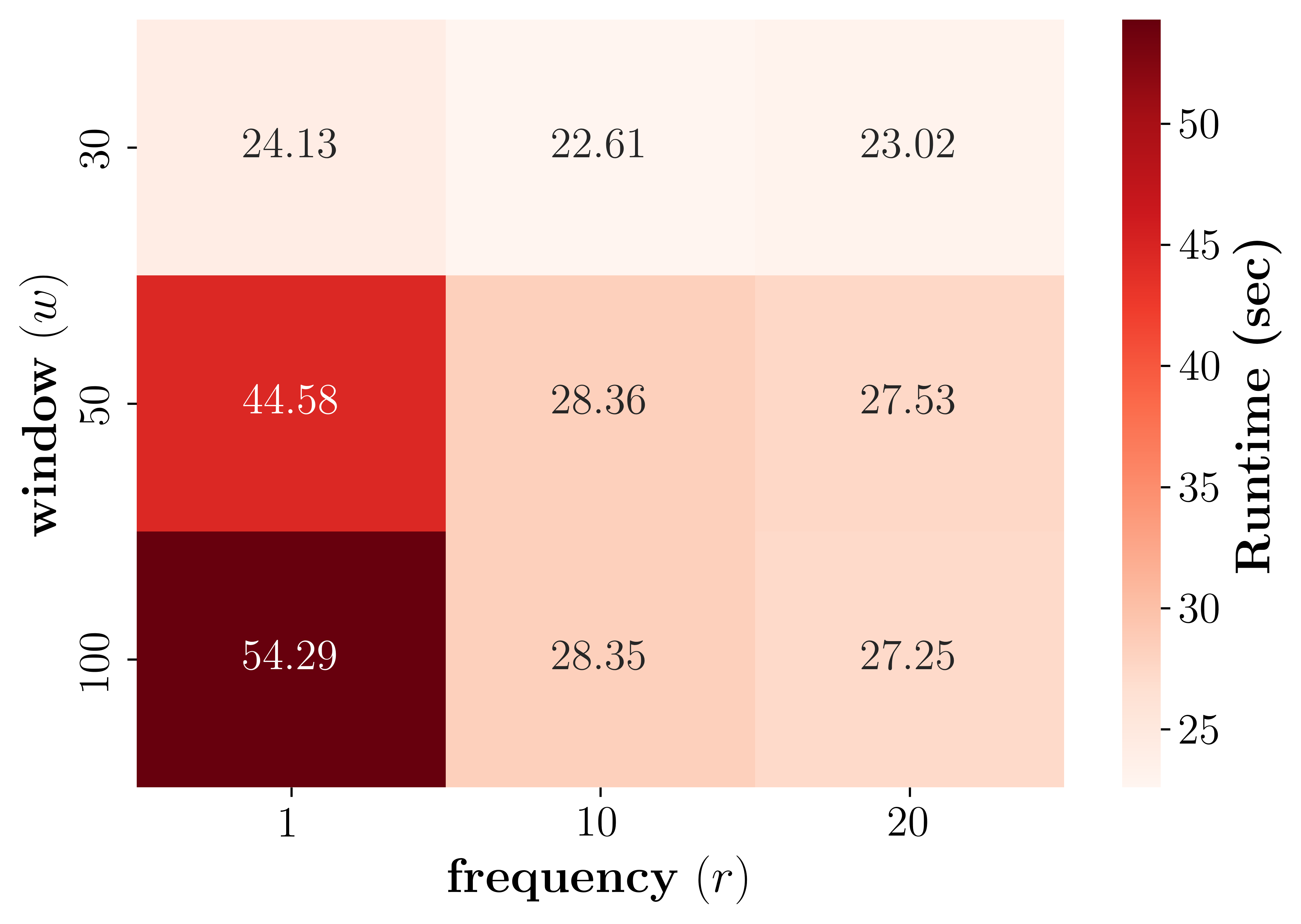}
    \end{minipage}
   \caption{
Sensitivity analysis of window size ($w$) and stride ($r$) in motion deblurring with DRUNet. \textbf{Left}: Peak PSNR for different $(w, r)$ combinations. \textbf{Right}: Runtime for 1000 iterations. The results show that choosing $w \in [30,40]$ and $r \in [10,20]$ gives a good tradeoff between reconstruction quality and computational efficiency, supporting our choice of parameters.
}

  \label{fig:ablation_wr}
\end{figure}

\begin{table}[t]
\caption{Peak and final PSNR (averaged over Set15C) for Vanilla RED and SKOOP-RED. The final PSNR is measured after $1000$ iterations, $\sigma$ is the standard deviation of the Gaussian blur kernel, and $\sigma_n$ is the noise level.\\
}
\vspace{0.4em}
\centering
\setlength{\tabcolsep}{2.5 pt}
\renewcommand{\arraystretch}{1.1}
\begin{tabular}{|>{\centering\arraybackslash}p{2.5cm}|c|cc|cc|cc|cc|}
\hline
\textbf{setting} & \textbf{method} & \multicolumn{2}{c|}{\textbf{DnCNN}} & \multicolumn{2}{c|}{\textbf{DRUNet}} & \multicolumn{2}{c|}{\textbf{GS-DRUNet}} & \multicolumn{2}{c|}{\textbf{DiffUNet}} \\
\cline{3-10}
& & peak & final & peak & final & peak & final & peak & final \\
\hline
\multicolumn{10}{|c|}{\textbf{Gaussian Deblurring}} \\
\hline
\multirow{2}{*}{\shortstack{$25{\times}25$ ($\sigma{=}1.0$)\\$\sigma_n{=}1/255$}} 
& Vanilla & 32.21 & 10.97 & 32.17 & 10.88 & 32.32 & 10.98 & 32.30 & 10.96 \\
& SKOOP & \textbf{33.56} & \textbf{33.56} & \textbf{33.78} & \textbf{33.78} & \textbf{34.14} & \textbf{34.14} & \textbf{34.00} & \textbf{34.00} \\
\hline
\multirow{2}{*}{\shortstack{$11{\times}11$ ($\sigma{=}1.6$)\\$\sigma_n{=}2.5/255$}}
& Vanilla & 27.80 & 11.05 & 27.71 & 10.94 & 27.78 & 11.07 & 27.68 & 10.96 \\
& SKOOP & \textbf{28.72} & \textbf{28.72} & \textbf{28.43} & \textbf{28.43} & \textbf{28.84} & \textbf{28.84} & \textbf{27.93} & \textbf{27.92} \\
\hline
\multirow{2}{*}{\shortstack{$15{\times}15$ ($\sigma{=}1.2$)\\$\sigma_n{=}5/255$}}
& Vanilla & 27.61 & 10.94 & 29.05 & 11.05 & 29.33 & 11.04 & 29.31 & 10.98 \\
& SKOOP & \textbf{27.63} & \textbf{27.63} & \textbf{29.48} & \textbf{29.48} & \textbf{30.03} & \textbf{30.03} & \textbf{29.54} & \textbf{29.54} \\
\hline
\multicolumn{10}{|c|}{\textbf{$2\times$ Superresolution}} \\
\hline
\multirow{2}{*}{\shortstack{$\sigma{=}1.2$\\$\sigma_n{=}0$}}
& Vanilla & 29.46 & 9.25 & 29.09 & 8.64 & 29.05 & 25.97 & 29.15 & 9.32 \\
& SKOOP & \textbf{29.61} & \textbf{29.61} & \textbf{29.09} & \textbf{29.09} & \textbf{29.05} & \textbf{29.05} & \textbf{29.38} & \textbf{29.37} \\
\hline
\multirow{2}{*}{\shortstack{$\sigma{=}1.0$\\$\sigma_n{=}1/255$}}
& Vanilla & 29.59 & 16.4 & 29.73 & 14.85 & 29.60 & 21.15 & 29.20 & 15.22 \\
& SKOOP & \textbf{29.59} & \textbf{29.59} & \textbf{29.73} & \textbf{29.73} & \textbf{29.67} & \textbf{29.67} & \textbf{29.25} & \textbf{29.25} \\
\hline
\end{tabular}
\label{tab:psnr_sr}
\end{table}

\subsection*{Practical guide: checkpointing and computational tradeoffs}
\begin{itemize}[leftmargin=*, itemsep=0pt]
    \item Koopman updates use only the $w$ most recent iterates, taken at regular intervals ($r$ iterations). This checkpointing strategy reduces both computational and memory costs compared to continuous updates.
    \item \textit{Sensitivity Analysis:} Small $r$ (frequent checks) improves instability detection but increases overhead; large $r$ reduces cost but can delay intervention. Larger $w$ increases noise robustness with little runtime impact.
    \item \textit{Recommended default:} $w=30$ and $r=10$ provide an effective balance between detection accuracy and computational efficiency (see Fig.~\ref{fig:ablation_wr}), with performance robust across a range of values.
   \item Parameters are user-tunable to suit different hardware or application needs. For example, using a higher $r$ (less frequent updates) reduces runtime overhead, which is useful for embedded or real-time imaging systems. In contrast, a lower $r$ (more frequent updates), especially when combined with a larger $w$ (longer history), enables finer-grained instability detection in applications where reconstruction reliability is critical, such as medical imaging or scientific analysis.
    \item In all tested scenarios, the added cost of stability monitoring (feature extraction and Koopman update at checkpoints) is negligible compared to the dominant cost of denoising and forward model computation.
\end{itemize}

\end{document}